\documentstyle[aps,multicol,epsf]{revtex}

\begin{document}
\draft
\title{Transport properties of heterogeneous materials
derived from Gaussian random fields: Bounds and Simulation.}
\author{A. P. Roberts}
\address{
Department of Applied Mathematics,
Research School of Physical Sciences,
Australian National University,
Canberra, ACT, 0200, Australia
}
\author{Max Teubner}
\address{
Max-Plank-Institut f\"ur Biophysikalische Chemie,
Postfach 2841, 3400 G\"ottingen, Federal Republic of Germany}
\date{\today}
\maketitle
\begin{abstract}
We investigate the effective conductivity ($\sigma_e$)
of a class of amorphous media defined by the level-cut
of a Gaussian random field.
The three point solid-solid correlation function is
derived and utilised in the evaluation of the
Beran-Milton bounds. Simulations are used to calculate
$\sigma_e$ for a variety of fields and volume fractions
at several different conductivity contrasts.
Relatively large differences in $\sigma_e$ are
observed between the Gaussian media and the identical
overlapping sphere model used previously as a `model'
amorphous medium. In contrast $\sigma_e$ shows
little variability between different Gaussian media.
\end{abstract}
\begin{multicols}{2}
\section{Introduction}
\label{introduction}
\begin{picture}(0,0)
\put (35,+292){ \parbox{16.0cm}{\small \raggedleft %
{\em Appeared in} PHYSICAL REVIEW E, VOL.\ 51, %
PAGES 4141$\;$-$\;$4154, May 1995}}
\end{picture}
The calculation of the effective transport properties of
random composite media is important in many scientific
and engineering applications\cite{TorqRev91}. 
Several techniques (effective medium
approximations and cluster expansions) have been developed for
predicting the effective properties of such materials
(briefly reviewed in ref.\ \cite{Thovert90}). 
However difficulties encountered in such methods have
provided the impetus for the development of rigorous bounds
\cite{Hashin62,Beran65a,Beran65b,McCoy,Milton81a,Milton81b}.
Such bounds rely on statistical descriptions of
the microstructure of the material which are available
for relatively few classes of media.
The advancement of computing technology
\cite{Kim90,Kim91,Kim92,Bonnecaze90} has also made
direct simulation of effective properties feasible.
It is the latter two approaches which we shall
discuss here, in the context of the effective conductivity
of a three dimensional amorphous isotropic two phase material.

There has been significant advances in the evaluation
of the Beran-Milton\cite{Beran65a,Milton81a} (BM) bounds
in the past decade.
The key parameter $\zeta_1$ (or $\zeta_2=1-\zeta_1$)
which incorporates microstructural information regarding the
composite has been evaluated primarily for materials comprised
of statistically independent cells \cite{Miller69,Milton82a}
or dispersions of regularly shaped inclusions
\cite{TorqStel83,Berryman85,Thovert90,Torquato85b}.
The simulation of the effective conductivity of continuum
random media is a computationally intensive process and
has only recently been studied for the second
class of materials\cite{Kim91,Kim92}.
Such an approach provides a basis for testing the
bounding theories and for generating outright predictions
of the effective properties of composites.

A class of materials which is not, in general, well
described by cellular or particulate models is
that of amorphous composites.
Such materials arise in certain alloys \cite{CorsonII,Li92},
microemulsions \cite{Teubner87,Pieruschka94} and
other systems \cite{Debye57}. 
The model which best captures some of the salient
features of such composites is the spatially uncorrelated
penetrable sphere (or the identical overlapping sphere)
model \cite{Weissberg63}.
Due to the simplicity in evaluating the statistical
correlation functions of such a material it has
served as a useful `model' amorphous
medium \cite{Doi76,TorqStel83,Stell87}.
However specific features of this model restrict its generality.
The inclusion (sphere)
phase and the matrix phase are topologically very
different, the small scale structure of the phase
boundaries is spherical and there are no long
range correlations in the model.
An alternative approach is to empirically measure
the specific correlation functions of a sample and
to apply the results in the evaluation of bounds
\cite{CorsonI,CorsonII,CorsonIII,CorsonIV}.
This approach is complicated and subject to error.
It is therefore interesting to seek a more complex model
of amorphous composites, yet simple enough so that
the correlation functions can be calculated.

Another method of modeling random composites is to define the
interface between the phases as a level cut of some random field
\cite{Cahn65,Skal73a,Skal73b,Zallen71,Blumenfeld93}
(see ref.\ \cite{Isichenko92} for a review).
Recent progress \cite{Berk87,Teubner91,Berk91} in the
theory of interfaces of level-cut Gaussian random fields
has made it possible to calculate the statistical information
necessary for the evaluation of the BM bounds.
There is evidence that the Gaussian random interface model
is a good approximation to certain oil-water
microemulsions\cite{Pieruschka92,Pieruschka93,Pieruschka94}
and we conjecture that it is a reasonable model
for amorphous alloys.

In this paper we investigate the effective conductivity
of such media using the above mentioned bounding techniques
and computer simulations.
The results are compared with the previously studied models
to demonstrate the differences that arise.
The paper is organised as follows.
In section \ref{bounds} we describe the equations governing
the electric field in a composite medium and the
bounds on the effective conductivity.
In sections \ref{correlation} \& \ref{determination}
we derive the statistical correlation functions for the
random media and apply them in the
calculation of the microstructure parameter.
Sections \ref{generation}-\ref{results} are
concerned with the generation of the random materials,
the simulation of the effective conductivity and comparison
of the data with the bounds. 
\section{Bounds on the effective properties of
composite materials}
\label{bounds}
The relationship between the current density $j$ and the
electric field $E=-\nabla\phi$ is given by Ohms Law,
\begin{equation} {\bf j}=-\sigma \nabla\phi. \end{equation}
Where, due to charge conservation, $\phi$ satisfies,
\begin{equation}  \nabla^2\phi=0 \end{equation}
throughout the material.
At the boundary of different regions of the material
with conductivities $\sigma_1$ and $\sigma_2$ we have,
\begin{equation}
\phi_1=\phi_2,\;\;\;\; \sigma_1\nabla\phi_1.{\bf n}=
\sigma_2\nabla\phi_2.{\bf n}
\end{equation}
The effective conductivity is defined by a macroscopic
form of Ohm's law,
\begin{equation}\label{effcon} \sigma_e=
\frac{<\sigma \nabla\phi>}{<\nabla\phi>}. \end{equation}
Now consider a composite material made up of two
components with conductivities $\sigma_1$ and $\sigma_2$
with volume fractions $p$ and $q=1-p$. The effective
conductivity will then depend on the $\sigma_i$, their
respective volume fractions, and the spatial distribution
(microstructure) of each phase \cite{Brown55}.

The first bounds on $\sigma_e$ were calculated by
Wiener \cite{Weiner12} who proved that
$<\sigma^{-1}>^{-1} \leq \sigma_e \leq <\sigma>$.
These bounds assume no details about the microstructure
and are hence valid for a general composite. 
As more statistical information regarding the composite is
included in the calculation of the bounds they become
more restrictive.
If the sample is assumed to be isotropic and macroscopically
homogeneous then the 2nd order bounds of
Hashin and Sthrikman\cite{Hashin62} are applicable.
To distinguish between such materials the third order
bounds of Beran are necessary.
(The term $n$th order bounds refers to the fact that
the bounds are exact to $O(\sigma_1-\sigma_2)^n$). 
The Beran \cite{Beran65a} bounds were derived using variational
principles and were subsequently simplified by Milton
\cite{Milton81a}.
Following the notation of Milton we define
$<a>=pa_1+qa_2$, $<\tilde{a}>=qa_1+pa_2$
(interchanging $p$ and $q$) and
$<a>_{\zeta}=\zeta_1a_1+\zeta_2a_2$.
Here $a_i=\sigma_i$ or $1/\sigma_i$.
In these terms the lower bound on $\sigma_e$ is,
\begin{equation}
\label{BMlo} \sigma_l= \left[ <\sigma^{-1}> -
\frac{2pq(\sigma_1^{-1}-\sigma_2^{-1})^2}
{2<\tilde{\sigma}^{-1}>+<\sigma^{-1}>_{\zeta}}  \right]^{-1} 
\end{equation}
while the upper bound is,
\begin{equation}  \label{BMhi} \sigma_u= \left[ <\sigma> -
\frac{pq(\sigma_1-\sigma_2)^2}
{<\tilde{\sigma}>+2 <\sigma>_{\zeta}}  \right]. 
\end{equation}
The so called microstructure parameter $\zeta_1$ is given
by a number of equivalent
integrals\cite{Beran65a,Milton81a,Milton82a}, of which
the formulation due to Brown \cite{Brown55}
is the best for our purposes,
\begin{eqnarray}
\label{zeta}
\zeta_1&=& \nonumber
\frac9{2pq}\int_0^\infty\!\!\frac{dr}{r} \int_0^\infty
\!\! \frac{ds}{s} \int_{-1}^1 du  P_2(u) \times \\
&&\left( p_{3}(r,s,t)-\frac{p_{2} (r) p_{2}(s)}{p}
\right) 
\end{eqnarray}
where $t^2=r^2+s^2-2rs u$ and $P_2(u)=(3 u^2-1)/2$ is the
Legendre polynomial of order 2.
The functions $p_{n}$ are $n$-point solid-solid correlation
functions (see section \ref{correlation}) where
the `solid' is phase 1 and the `void' is phase 2. 

As Milton notes these bounds converge when $\zeta_1=0,1$
and are equal to one of the 2nd order Hashin-Sthrikman
bounds in each case.
An improved lower bound has been derived by
Milton \cite{Milton81b} for the case $\sigma_2>\sigma_1$.
In later sections we consider materials with
$\sigma_1>\sigma_2$ for which this bound is
(see ref.\ \cite{Kim92}), by interchanging the roles
of the materials,
\begin{equation}  \label{Mlo}
\sigma_l=\sigma_2 \frac{1+(1+2p)\beta_{12}-
2(q\zeta_1-p)\beta_{12}^2}
{1+q\beta_{12}-(2q\zeta_1+p)\beta_{12}^2},
\end{equation}
where $\beta_{12}=(\sigma_1-\sigma_2)/(\sigma_1+2\sigma_2)$. 
By way of mathematical analogy these bounds also apply to
to the effective dielectric, diffusion and magnetic permeability
coefficients of composite materials.
\section{Correlation Functions for the
Gaussian Random Interface model}
\label{correlation}
There is an extensive literature on the calculation of statistical
correlation functions \cite{TorqRev91}.
The case of the three point solid-solid correlation
function has been considered
empirically \cite{CorsonI,CorsonII,Berryman88,Lu90},
and theoretically for cellular materials \cite{Miller69,Brown74}
and spherical inclusions
\cite{Weissberg63,TorqStel83,Berryman85,TorqStel85} to name a few.
Here we take the interface between the phases to be defined
by a level cut of a random field \cite{Cahn65,Skal73b}.
Now consider a Gaussian random field
$y({\bf r})$ \cite{Wang45,Isichenko92}
(see section \ref{generation})
and let the level sets $y({\bf r})=\alpha$ define the interface
(with the region $y>\alpha$ being phase 1).
Then the $n$ point correlation function is given by
the volume average,
\begin{equation}\label{defnpn}
p_n({\bf r}_1,{\bf r}_2,\dots,{\bf r}_n)=
<H(y_1-\alpha)\dots H(y_n-\alpha)>
\end{equation}
where $H(y)$ is the Heavyside function and $y_i=y({\bf r}_i)$.
$p_n$ is then the probability that the $n$
points will lie in phase 1.
For a macroscopically homogeneous isotropic material
$p_n$ only depends on the distances
$r_{ij}=|{\bf r}_i-{\bf r}_j|$ between the points.
Since volume and ensemble averages are equivalent in
such a medium \cite{Brown55} we can use the latter to
evaluate eqn.\ (\ref{defnpn}).
The joint probability density of $y_i$ is,
\begin{equation}
P_n(y_1y_2\dots y_n)=\frac{1}{\sqrt{ (2\pi)^ n |G|} }
\exp( -\frac 12 {\bf y}^T G^{-1} {\bf y} ), \end{equation}
where the elements of $G$ are $g_{ij}=<y_iy_j>$ \cite{Wang45}.
The latter quantity we refer to
as the field-field correlation function,
\begin{equation}
g_{ij}= \int_0^{\infty}{4\pi k^2 \rho(k) \frac{\sin
k|{\bf r}_i-{\bf r}_j|}{k|{\bf r}_i-{\bf r}_j|}}{dk}.
\label{gdef}
\end{equation}
where $\rho(k)$ is the spectral density of the field.

Berk \cite{Berk87,Berk91} and Teubner \cite{Teubner91} have
derived the one point function (volume fraction),
\begin{equation}  \label{onepoint}
p= \frac{1}{\sqrt{2\pi}} \int_\alpha^\infty
\exp(-\frac12 t^2) dt \end{equation}
and the two point function,
\begin{equation}  \label{twopoint} p_2(g_{ij})= \frac1{2\pi}
\int_{0}^{g_{ij}}{ \exp{\left(-\frac{\alpha^2}{1+t}\right)}
{\frac{dt}{\sqrt{1-t^2}} }}+p^2 .  \end{equation}
The three point function is calculated using the
techniques described in ref.\ \cite{Teubner91}.
The following identities are used \cite{Wang45},
\begin{equation}
< \exp(i {\bf y}.{\bf w} ) >= \exp( -\frac12 {\bf w}^T G {\bf w} )
\end{equation}
and
\begin{equation}
H(y-\alpha)=\frac{-1}{2\pi i}\int_C e^{-iw(y-\alpha)}\frac{dw}{w}
\end{equation}
where the contour $C$ is directed along the real axis except
near the origin where it crosses the imaginary axis
in the upper half plane.
This leads (after algebra) to,
\begin{equation}
\frac{\partial p_3^T}{\partial g_{12}}=\frac{\partial p_2(g_{12})}
{\partial g_{12}} \frac 1{\sqrt{2\pi}}
\int^{\alpha}_{\alpha F_{12} } \exp(-\frac12 t^2) dt,
\end{equation}
where, 
\begin{equation}  F_{12}=\frac{\sqrt{1-g_{12}}}{\sqrt{1+g_{12}}}
\frac{1+g_{12}-g_{13}-g_{23}}{\sqrt{|G|} },
\end{equation}
and $|G|=1-g_{12}^2-g_{13}^2-g_{23}^2+2g_{12}g_{13}g_{23}$.
Similar expressions can be derived for
${ \partial p_3^T }/{ \partial g_{13} }$ and
${ \partial p_3^T }/{ \partial g_{23} }$.
Defining $A_{ij}={ \partial p_3^T }/{ \partial g_{ij} }$
we have therefore,
\begin{eqnarray}
p_3^T(g_{12},g_{13},g_{23})
     &=&g_{12} \int_{0}^{1} A_{12}(tg_{12},tg_{13},tg_{23})
dt \nonumber \\
     &+&g_{13} \int_{0}^{1} A_{13}(tg_{12},tg_{13},tg_{23})
dt \nonumber\\
     &+&g_{23} \int_{0}^{1} A_{23}(tg_{12},tg_{13},tg_{23}) dt. 
\end{eqnarray}
The truncated three point correlation function $p_3^T$
is related to the $p_3$ by the expression,
\begin{eqnarray}
\label{threepoint}
p_3(g_{12},g_{13},g_{23})&=&
p_3^T(g_{12},g_{13},g_{23})+p p_2(g_{12})
\nonumber\\ &+&p p_2(g_{13})+p p_2(g_{23})-2p^3.
\end{eqnarray}
To examine the limit $r_{12}\to0$ $(g_{12}\to1, g_{23}\to g_{13})$
set $f(g_{13})= p_3^T(1,g_{13},g_{13})$ then
$df(g_{13})/dg_{13}=(1-2p)dp_2(g_{13})/dg_{13}$
and $f(0)=0$ implying
$p_3^T(1,g_{13},g_{13})=(1-2p)(p_2(g_{13})-p^2)$
as it should. (Similarly in the other limits).
The X-ray spectra of these materials can be calculated
from $p_2$ \cite{Debye57,Teubner87,Berk91} and hence they
can be related to physical composites.
Furthermore it has been shown that the
surface to volume ratio is given by \cite{Debye57,Teubner91},
\begin{equation}
\frac SV=\frac 2\pi e^{-\frac12 {\alpha^2} } \sqrt{\frac13 <k^2>}.
\end{equation}

As the evaluation of the integrals in eqns.\ (\ref{twopoint}) \&
(\ref{threepoint}) are computationally intensive it is
useful to derive various approximations.
Rigorous approximations for $p_2$ for the cases 
$|\alpha|\ll1$ and $|\alpha|\gg 1$ are derived in
appendix~\ref{asymptotic}.
A useful non-rigorous approximation to $p_{123}^T$
can be developed by requiring that the approximation
have similar properties to the actual function for
$r_{ij}\gg1$ and satisfy the known consistency conditions in
various limits \cite{Brown55}. 
Using the compact notation $p_{ij}^T=p_2(g_{ij})-p^2$ and
$p_{ijk}^T=p_3^T(g_{ij},g_{ik},g_{jk})$ we have ($r_{ij}\gg1$),
\begin{eqnarray}
p_{ij}^T &\propto& g_{ij} \\
p_{123}^T&\propto&g_{12}g_{13}+g_{12}g_{23}+g_{13}g_{23}. 
\end{eqnarray}
Using this information, and including a higher order term for
consistency ($p_3(r_{12},r_{12},0)=p_2(r_{12})$) we construct,
\begin{eqnarray}
\label{nonrig}
p_{123}^T & \approx & \frac{1-2p}{2p(1-p)}\left( \nonumber
p^T_{12}p^T_{13}+p^T_{12}p^T_{23}+p^T_{13}p^T_{23}\right) \\
&& - \frac{1-2p}{2p^2(1-p)^2} p^T_{12}p^T_{13}p^T_{23}.
\end{eqnarray}
We note that this approximation has a maximum absolute error of 
$O(10^{-3})$ for the materials considered here.
As such it is an
order of magnitude better than a previously suggested
approximation \cite{Weissberg62}
$p_3(r_{12},r_{13},r_{23})\approx p_2(u)p_2(v)/p$ where
$u$ and $v$ are the smallest, and next to smallest, values of
$r_{12}, r_{13}$ and $r_{23}$.
\section{Determination of $\zeta_1$}
\label{determination}
Actual calculations of the microstructural parameter
$\zeta_1$ (eqn.\ (\ref{zeta})) have, to date, been for
four classes of materials.
Cellular materials \cite{Miller69,Milton82a}, empirically
measured physical composites \cite{CorsonIV}, periodic
arrays of spheres \cite{McPhedran81} and materials
with spherical inclusions.
In the latter class the cases studied include:
identical overlapping spheres \cite{TorqStel83,Berryman85}
(the IOS model), identical hard spheres \cite{TorqLado86}, 
and poly-dispersed spheres \cite{Stell87,Thovert90} (many
of these results are summarised in ref.\ \cite{TorqRev91}). 

We now describe aspects of the computation of $\zeta_1$ for
several spectra of the Gaussian random interface (GRI) model. 
It can be shown that $\zeta_1=\frac12$ for $p=\frac12$ 
\cite{Miller69,Brown74,Milton82a} (see appendix \ref{zhalf})
and that $\zeta_1=1-\zeta_2$ where $\zeta_2$ is the
microstructure parameter associated with phase 2.
As $\zeta_1$ is dimensionless it must depend only
on the ratios of the length scales associated with the
spatial variables in the (dimensionless) correlation functions.  
That this should be so also follows from a simple
dimensional analysis of the equations governing the electric
field (no physical length scale is present).
Henceforth  we scale all spatial variables against a
characteristic decay length without loss of generality. 

We consider three types of media generated from Gaussian
random fields.
The field-field correlation functions and their
corresponding spectra are: \\
Model I:
\begin{eqnarray}
\label{moda}
g(r)&=&e^{-r}\frac{\sin\nu r}{\nu r} \\
\rho(k)&=&\pi^{-2}\left( (1-\nu^2+k^2)^2+4\nu^2 \right)^{-1}.
\end{eqnarray}
Here $\nu=2\pi l_1/ l_2$ with $l_1$ the decay length of the field
and $l_2$ the characteristic domain size.
When $r\gg0$ the correlation functions arising from
this model are similar to those considered in
refs. \cite{Debye57} ($\nu=0$) and \cite{Teubner87} ($\nu>0$).
Note that this model has an infinite surface to volume ratio
since $<k^2>$ diverges, however for computational realizations of
the model the ratio is finite (see section \ref{generation})
and the model is well defined.
However it is interesting to study $\zeta_1$ for this model
to investigate the effect of interfacial roughness on effective 
properties.  \\
Model II:
\begin{eqnarray}
g(r)&=&e^{-r^2} \\
\rho(k)&=&\frac{e^{-\frac14{k^2}}}{(4\pi)^\frac32}
\end{eqnarray}
\\
Model III:
\begin{eqnarray}
g(r)&=& \frac
{3\left(\sin \mu r-\mu r\cos \mu r-\sin r-r\cos r \right)}
{r^3(\mu^3-1)} \\
\rho(k)&=&\frac{3}{4\pi(\mu^3-1)} \left( H(\mu)-H(1) \right)
\label{modf}
\end{eqnarray}
where $\mu=k_1/k_0$.
Note that $\rho(k)\to \delta(k)$ as $\mu\to 1$
and the simple model used by Berk \cite{Berk87} is recovered.

To perform the integration (\ref{zeta}) we use cylindrical
co-ordinates (which damp the singularity at the origin),
interchange the order of integration and exploit the
$r-s$ symmetry to give,
\begin{equation}
\zeta_1=\frac{9}{2pq}
\int_{-1}^{1} \left( \int_0^\infty \int_0^{\frac\pi4}
I(w,\phi,u) dw d\phi \right) P_2(u) du 
\end{equation}
with
\begin{equation}  I(w,\phi,u)=
\frac{p\;p_3(r,s,t)- p_2(r)p_2(s)}{p^3 w\sin\phi\cos\phi}.
\end{equation}
To elucidate the nature of the singularity in the integrand at
$w=\phi=0$ we consider $w$, $\phi$ (and hence $r$, $s$ and $t$)
to be small and assume the form $g(r)\approx 1-a r^n$
(where $n=1,2$ in accord with models I-III).
Now for $p=\frac12$ the numerator of the expression for
$I$ is given by,
\begin{equation}
\frac{\sin^{\!-\!1}(g(t))}{8\pi}-
\frac{\sin^{\!-\!1}(g(r))\sin^{\!-\!1}(g(s))}
{4\pi^2}\approx\sqrt{2aw^n\phi^n},
\end{equation}
where we have used the results
$\sin^{-1}(g(r))\approx\frac\pi2-\sqrt{2ar^n}$, $r=w\cos\phi$,
$s=w\sin\phi$ and $t=w\sqrt{1-\sin2\phi u}$. Therefore 
$I(w,\phi,u) \sim (w\phi)^{\frac n2-1}$ with $p=\frac12$
and numerical analysis shows this scaling also
holds for $p\ne\frac12$.
An integration rule which takes the singularity into
account is employed. Note that for finite surface area to volume
ratios $n=2$.
The number of abscissae in each of three integration ranges
was increased until the third significant figure in the
estimation of $\zeta_1$ remained constant. The integration
method was tested on the known correlation functions for the IOS
model \cite{Weissberg63,TorqStel83}.
The results are in exact agreement (to the reported
3rd significant figure) with those of Torquato and Stell
\cite{TorqRev91,TorqStel83} and agree to the second
significant figure with the results of
Berryman \cite{Berryman85}.
The calculation of the
correlation functions in the integrand is done
using a combination of iterative quadrature
rules\cite{Press}, the asymptotic results presented in appendix
\ref{asymptotic} and the non-rigorous
approximation for the truncated three point function (\ref{nonrig}).
The latter is used whenever $1-g<10^{-5}$ since
in this case the functions $F_{ij}$ exhibit
large derivatives and the quadrature rules converge too slowly.
The accuracy sought in the application of each
approximation is $O(10^{-6})$.

\noindent
\begin{minipage}[b!]{8.5cm}
\begin{table}
\caption{The microstructure parameter $\zeta_1$ for
various GRI models.  Prior results for the identical
overlapping sphere (IOS) model 
\protect\cite{TorqRev91,TorqStel83} and the symmetric
spherical cell (SSC) model \protect\cite{Miller69,Milton81a}
are included for purposes of comparison.}
\label{tabzeta1}
\begin{tabular}{ccccccc}
\multicolumn{1}{c}{ } &
\multicolumn{2}{c}{Prior Results}   &
\multicolumn{4}{c}{GRI model}   \\
\hline
p & SSC & IOS & I $\nu=0$ & I $\nu=10$ & II & III $\mu=1.5$ \\
\tableline
0.01 & 0.01 &       & 0.269 & 0.193 & 0.099 & 0.060 \\
0.05 & 0.05 &       & 0.293 & 0.217 & 0.160 & 0.105 \\
0.1  & 0.1  & 0.056 & 0.319 & 0.248 & 0.210 & 0.150 \\
0.2  & 0.2  & 0.114 & 0.366 & 0.309 & 0.291 & 0.237 \\
0.3  & 0.3  & 0.171 & 0.411 & 0.372 & 0.363 & 0.324 \\
0.4  & 0.4  & 0.230 & 0.456 & 0.436 & 0.432 & 0.411 \\
0.5  & 0.5  & 0.290 & 0.500 & 0.500 & 0.500 & 0.500 \\
0.6  & 0.6  & 0.351 & 0.544 & 0.564 & 0.568 & 0.588 \\
0.7  & 0.7  & 0.415 & 0.589 & 0.628 & 0.636 & 0.675 \\
0.8  & 0.8  & 0.483 & 0.634 & 0.691 & 0.709 & 0.763 \\
0.9  & 0.9  & 0.558 & 0.681 & 0.752 & 0.790 & 0.850 \\
0.95 & 0.95 & 0.604 & 0.707 & 0.783 & 0.840 & 0.895 \\
0.99 & 0.99 & 0.658 & 0.731 & 0.807 & 0.901 & 0.940       
\end{tabular}
\end{table}
\end{minipage}

\noindent
\begin{minipage}[b!]{8.5cm}
\begin{figure}
\centering \epsfxsize=8.5cm\epsfbox{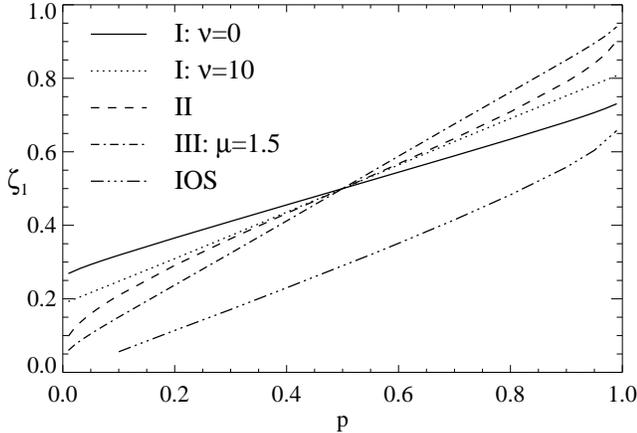}

\vspace{3mm}

\caption{
\label{figzeta}
The microstructure parameter $\zeta_1$ vs. $p$ for the GRI
models (equations (\protect\ref{moda})-(\protect\ref{modf}))
and the IOS model \protect\cite{TorqRev91,TorqStel83}.}
\end{figure}
\end{minipage}

\vspace{3mm}

The results for each of the models is presented in
table \ref{tabzeta1} and plotted in figure \ref{figzeta}
along side the results for the IOS model.
Several comments on the qualitative relationship between
$\zeta_1$ for the GRI model and prior calculations can be made. 
Consider the expansion,
\begin{equation}\zeta_1=\sum_{i=0}^{\infty} e_i p^i.\end{equation}
For a general class of materials with spherical inclusions
it has been shown that $e_0=0$ \cite{Torquato85b,Thovert90}.
If the inverse of these materials is considered
(so that the correlation functions refer to the
material surrounding the inclusion) or ellipsoidal inclusions are
considered \cite{Torquato85b} then $e_0>0$ (see appendix
\ref{qualitative}).
For the case of symmetric cell materials
$e_0=M\!\in\![0,1]$ where $M=0$ for spherical cells and $M=1$
for plate like cells \cite{Milton81a,Miller69}.
Another interesting feature of $\zeta_1$ is that it
is observed to be linear with $p$ over a wide range
\cite{TorqRev91,Thovert90}.
Indeed for the symmetric cell model $e_1=(1-2M)$ and
$e_2,e_3\dots=0$.
By inspection of figure \ref{figzeta} we see that
$e_0>0$ in qualitative agreement with the results
for non-spherical inclusions which will occur
in the GRI models.
Also note that $\zeta_1$ is very similar to the results
for the symmetric cell model for some $0<M<1$ over a
wide range of $p$.
This discussion demonstrates the success of
simple models of random media to capture the
qualitative features of $\zeta_1$ for the
amorphous materials considered here.

Calculations of the related microstructure parameter
$\eta_1$ which arises in bounds on the elastic properties
of random composites are reported in appendix \ref{elastic}.
\section{Generation of Fields}
\label{generation}
For computational purposes we consider a $T$-periodic
Gaussian random field \cite{Wang45} with a
maximum wavenumber $K=2\pi N/T$,
\begin{equation}  \label{yK}
y_{\!\!\!\mbox{ {\tiny $K$} }\!\!\!} ({\bf r})=
\sum_{l=-N}^{N}
\sum_{m=-N}^{N}
\sum_{n=-N}^{N} c_{lmn} e^{i{\bf k}_{lmn}.{\bf r}}
\end{equation}
where, 
\begin{equation}
{\bf k}_{lmn}=\frac{2\pi}{T}(l{\bf i}+m{\bf j}+n{\bf k}).
\end{equation}
For $y_{\!\!\!\mbox{ {\tiny $K$} }\!\!\!} $ real we
require $c_{l,m,n}=\bar c_{-l,-m,-n}$
and as $<y_{\!\!\!\mbox{ {\tiny $K$} }\!\!\!} >=0$
we set $c_{0,0,0}=0$.
For reasons which become clear below we
also take $c_{lmn}=0$ for $k_{lmn}=|{\bf k}_{lmn}|\ge K$. 
With $c_{lmn}=a_{lmn}+ib_{lmn}$, $a_{lmn}$ and $b_{lmn}$ are
random independent variables (subject to the
conditions on $c_{lmn}$) with Gaussian
distributions such that $<\!a_{lmn}\!>=<\!b_{lmn}\!>=0$ and
\begin{equation}  \label{rho}
<a_{lmn}^2>=<b_{lmn}^2>=
\frac12\rho_{\!\!\!\mbox{ {\tiny $K$} }\!\!\!}
(k_{lmn})\left({\frac{2\pi}{T}}\right)^3
\end{equation}
with $\rho_{\!\!\!\mbox{ {\tiny $K$} }\!\!\!} (k)$
the spectral density.
The field-field correlation function $g_{\!\!\!\mbox{ {\tiny $K$}
}\!\!\!}$ is given by
\begin{eqnarray}
g_{\!\!\!\mbox{ {\tiny $K$} }\!\!\!}(r_{12})&=&
<y_{\!\!\!\mbox{ {\tiny $K$} }\!\!\!}
({\bf r}_1)\overline{y_{\!\!\!\mbox{ {\tiny $K$} }\!\!\!} ({\bf r}_2)}>
\\ &=&
\sum_{-N}^{N}
\sum_{-N}^{N}
\sum_{-N}^{N}\!\!
<c_{lmn}\overline{c_{lmn}}> e^{i{\bf k}_{lmn}.
({\bf r}_1-{\bf r}_2)}
\\&\approx& \int_0^{K}{4\pi k^2
\rho_{\!\!\!\mbox{ {\tiny $K$} }\!\!\!} (k) \frac{\sin
k|{\bf r}_1-{\bf r}_2|}{k|{\bf r}_1-{\bf r}_2|}}{dk}.
\label{gkdef}
\end{eqnarray}
The last integral is obtained by taking $N$ and $T$ large, using
equation (\ref{rho}) and recognising that the summation is the
approximation of a triple integral.  
Following a transformation to spherical co-ordinates
we integrate over the angular variables to obtain (\ref{gkdef}).
Since $g_{\!\!\!\mbox{ {\tiny $K$} }\!\!\!} (0)=1$ we define,
\begin{equation}
\rho_{\!\!\!\mbox{ {\tiny $K$} }\!\!\!} (k)=P^{-1} \rho(k),\;\;
P=\int_0^K 4\pi k^2 \rho(k) dk \end{equation}
where the $\rho(k)$ are defined in section \ref{correlation}. 
If in addition we take $K\to\infty$ then the conventional
correlation function (and spectral density) are recovered. 
The Fourier expansion (\ref{yK}) is evaluated using a FFT. 

Consider materials derived from the field
$y_{\!\!\!\mbox{ {\tiny $K$} }\!\!\!} $ (eqn.\ (\ref{yK}))
as discussed in section \ref{correlation}.
Cross sections of the media for four different variants of
the models are plotted in figure \ref{figmodels} (a)-(d). 
The large scale structure of the interface is determined
by the terms in the expansion with small $k$
while the small scale structure (ripples on the surface)
are determined by the terms where $k$ is large.
A physical material will naturally contain a finite
cutoff wavelength either imposed by the molecular size
or by the manifestation of surface tension at
the phase boundaries.
This wavelength will then dictate the grid resolution
necessary to properly resolve the structure for simulational
purposes.
Conversely for this study the discretisation
$\Delta x= T/M$, where $M$ is the number of grid points
in the $x$ direction, will restrict the choice of $K$.
Thus $\Delta x \ll \lambda_{min}= 2\pi/K$ or $N\ll M$. 
Another constraint

\end{multicols}

\begin{center}
\noindent
\begin{minipage}[b!]{15cm}
\begin{figure}
\centering \epsfxsize=14.0cm\epsfbox{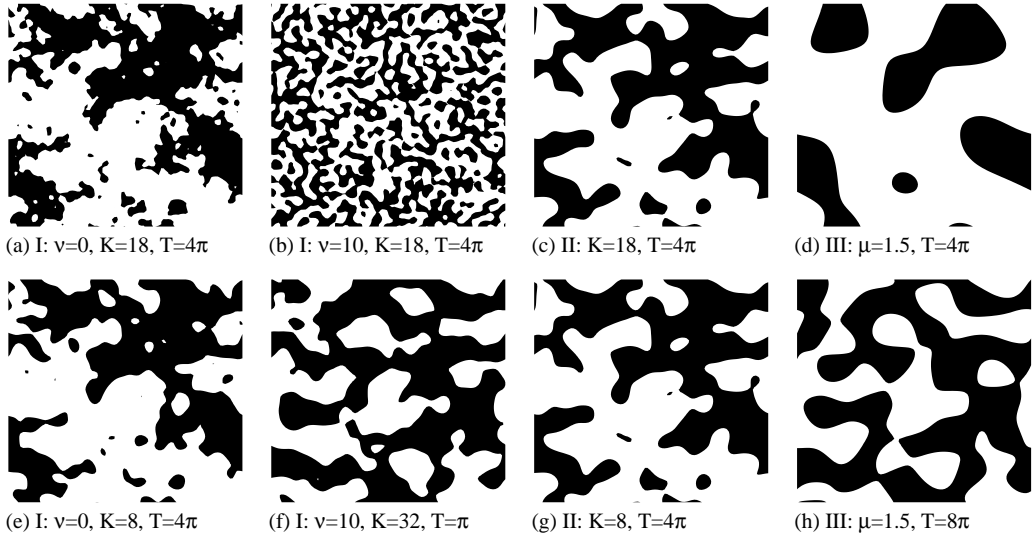}

\caption{
\label{figmodels}
Cross sections of the models generated at the same
scale (a)-(d) and at the scale to be used in the
simulations (e)-(h).
The volume fraction $p=0.5$
and each of the fields are generated using the same
random number seed to clearly show the differences amongst
the models.
The parameters
$\nu$, $\mu$, $T$ and $K$ are discussed in the text.}
\end{figure}
\end{minipage}
\end{center}

\begin{multicols}{2}

\noindent
is that $T\gg l_d$ where $l_d$ is an
effective decay length of the field defined by $g(l_d)=e^{-1}$.
This must be so to ensure that the edges of
the sample are uncorrelated to properly
simulate an infinite random medium.
For the approximation involved in determining
$g_{\!\!\!\mbox{ {\tiny $K$} }\!\!\!} $
(eqn.\ (\ref{gkdef})) to be accurate would
require $T\gg 2\pi$, however in this study this constraint
was found to be of less importance and is ignored.

The computational parameters used for each of
the four variants of the models are given in table
\ref{tabcomp}.
Cross-sections of each of the materials are plotted
in figures \ref{figmodels} (e)-(h) and the interface for
models II and III are plotted in figures
\ref{figIIp.5} and \ref{figIIIp.5}.
To properly account for the effect of $K$ for the computational
models $\zeta_1$ must be recalculated using
$g_{\!\!\!\mbox{ {\tiny $K$} }\!\!\!} (r)$.
This is done using a look up table generated by
numerical integration of (\ref{gkdef}) and an
asymptotic expansion for large $r$
(where the integral is costly to evaluate),
\begin{equation}  \label{gkasy}
g_{\!\!\!\mbox{ {\tiny $K$} }\!\!\!} (r)=
\frac1{P}g(r)-4\pi K\rho_{\!\!\!\mbox{ {\tiny $K$} }\!\!\!} (K)
\cos Kr\frac{1}{r^2}+ O\left( {\frac1{r^3}} \right). 
\end{equation}

The function $g_{\!\!\!\mbox{ {\tiny $K$} }\!\!\!} (r)$
for model I is plotted along with
direct measurements of $<y({\bf r}_1)y({\bf r}_2)>$ in
figure \ref{figgK8}.
The agreement is seen to very good.
The values of $\zeta_1$ for the computational models are
presented in table \ref{tabzeta2}.
The effect of $K$ on $\zeta_1$ can be seen
most clearly for Model I where
the surface area to volume ratio becomes arbitrarily large as
$K$ increases.  The interfacial smoothing effect of
imposing a finite cut-off ($K$) is shown in
figures \ref{figmodels} (a) ($K=18$) and (e) ($K=8$).

\noindent
\begin{minipage}[b!]{8.5cm}
\begin{table}
\caption{The parameters used to generate the computational
materials discussed in sections \protect\ref{generation} \&
\protect\ref{results}.}
\label{tabcomp}
\begin{tabular}{ccccc}
\multicolumn{2}{c}{Model} &
\multicolumn{1}{c}{T}   &
\multicolumn{1}{c}{N}   &
\multicolumn{1}{c}{K}   \\
\tableline
I  & $\nu=0$   & $4\pi$ & 16 & 8  \\
I  & $\nu=10$  & $\pi$  & 16 & 32 \\
II & $-$       & $4\pi$ & 16 & 8  \\
III& $\mu=1.5$ & $8\pi$ & 6  & 1.5
\end{tabular}
\end{table}
\end{minipage}

\noindent
\begin{minipage}[b!]{8.5cm}
\begin{figure}
\centering \epsfxsize=7.0cm\epsfbox{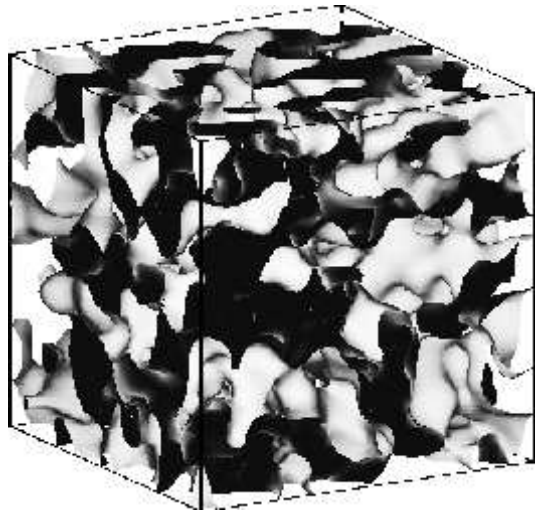}

\vspace{3mm}

\caption{
\label{figIIp.5}
A plot of the interface
$y_{\!\!\!\mbox{ {\protect\tiny $K$} }\!\!\!}({\bf r})=0$
($p=0.5$) for Model II. The parameters
used to generate the field are $K=8,T=4\pi$.
Note that the large scale structure of this model
is similar to that of model I ($\nu=0$) as can be seen
by comparing figs.\ \protect\ref{figmodels} (e) \& (g).}
\end{figure}
\end{minipage}

\noindent
\begin{minipage}[b!]{8.5cm}
\begin{figure}
\centering \epsfxsize=7.0cm\epsfbox{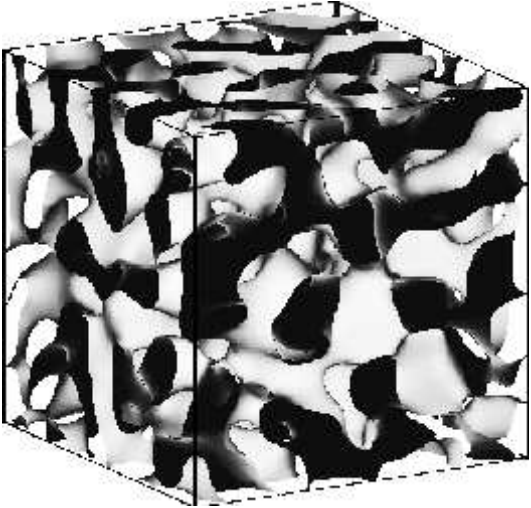}

\vspace{2mm}

\caption{
\label{figIIIp.5}
A plot of the interface $y_{\!\!\!\mbox{ {\protect\tiny $K$} }\!\!\!}
({\bf r})=0$ ($p=0.5$) for Model III. The parameters used to
generate the field are $\mu=1.5,T=8\pi$.}
\end{figure}
\end{minipage}

\vspace{2mm}

This has the effect of decreasing (increasing) the magnitude
of $\zeta_1$ for $p<\frac12$ ($p>\frac12$) (compare tables
\ref{tabzeta1} and \ref{tabzeta2}). For small (and
large) volume fractions this can be qualitatively explained
by the fact that the inclusion phase of the model will be much
rougher (and hence less `spherical') for $K=\infty$ than for $K=8$
(See appendix \ref{qualitative}).
$\zeta_1$ for model II is unchanged for finite $K$ to
the accuracy calculated here.

\vspace{-8mm}

\section{Solution of the Laplace equation}
\label{solution}

\vspace{-6mm}

\noindent
There are a variety of different methods of simulating
the effective conductivity of an inhomogeneous medium.
These include direct solution of the partial differential
equations governing the potential $\phi$ \cite{Saeger91},
Brownian motion algorithms\cite{Kim90,Kim91,Kim92}, Fourier 
methods \cite{Shen90} and other techniques\cite{Bonnecaze90}.

\noindent
\begin{minipage}[b!]{8.5cm}
\begin{figure}
\centering \epsfxsize=8.0cm\epsfbox{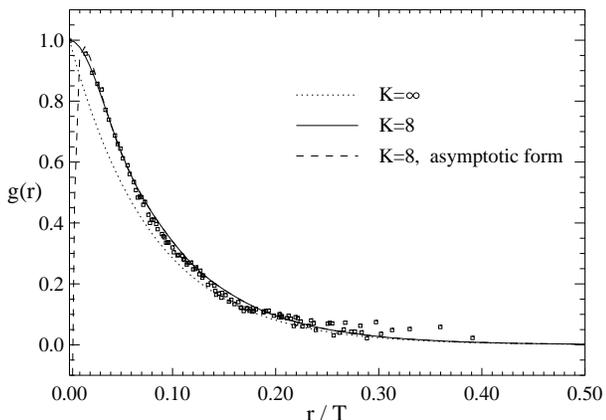}
\caption{
\label{figgK8}
The computational field-field correlation function
$g_\protect{\!\!\!\mbox{ {\protect\tiny $K$} }\!\!\!} (r)$
(eqn.\ (\protect\ref{gkdef}))
for model I $(\nu=0,K=8)$.
The exact form (ie.  $K=\infty$) given
by eqn.\ (\protect\ref{moda}) and the asymptotic form,
eqn.\ (\protect\ref{gkasy}), are included for purposes
of comparison. The data was measured from a single sample.}
\end{figure}
\end{minipage}

\noindent
\begin{minipage}[b!]{8.5cm}
\begin{table}
\caption{The microstructure parameter $\zeta_1$ for the 
computational materials listed in table \protect\ref{tabcomp}.
The parameter $K$ is the maximum wave number in
the Fourier Series. The relation $\zeta_1(p)=1-\zeta_1(1-p)$
can be used to determine $\zeta_1$ for $p\geq 1/2$.}
\label{tabzeta2}
\begin{tabular}{cccc}
p & I $\nu=0$  & I $\nu=10$  & II  \\
  & $K=8$ & $K=32$ & $K=8$ \\
\tableline
0.01 & 0.155 & 0.085 & 0.099  \\
0.05 & 0.214 & 0.136 & 0.160  \\
0.1  & 0.258 & 0.182 & 0.210  \\
0.2  & 0.326 & 0.265 & 0.291  \\
0.3  & 0.387 & 0.344 & 0.364  \\
0.4  & 0.444 & 0.422 & 0.432  
\end{tabular}
\end{table}
\end{minipage}

We solve Laplace's equation with the charge conservation
boundary conditions discussed in section \ref{bounds}
in a cube of side length $T$ using $M^3$ nodes.
A potential of $\phi_1$ and $\phi_0$ are applied on the faces
$z=0$ and $z=T$ and periodic boundary conditions are imposed
on the four faces parallel to the direction of the current
to model a sample of infinite extent in the $x$ and $y$ directions.
A finite difference scheme is used to approximate the
field and is solved using a conjugate gradient method.
In appendix \ref{implementation} we discuss the efficient
implementation of the algorithm on a parallel computer.
The $z$ components of the current and the field are then used
in equation (\ref{effcon}) to determine $(\sigma_e)_M$.
The convergence criterion of the CG solver is
decreased until the third significant figure of
$(\sigma_e)_M$ remained constant.
To estimate the continuum value of $\sigma_e$ we assume that
$\sigma_e\approx(\sigma_e)_M+a_1 M^{-1}$ and fit a line
(using least squares) to several values of
$(\sigma_e)_M$ vs. $M^{-1}$.
The intercept of this line with the axis $M^{-1}=0$ then
provides $\sigma_e$.

Before proceeding to the random media we simulate the effective
conductivity of a regular array of spheres 
of conductivity $\sigma_1=10$ in a matrix of
conductivity $\sigma_2=1$.
Exact results for this model have been calculated
by McKenzie {\it et al.} \cite{McKenzie78} and the model
has been used by previous authors \cite{Kim91,Bonnecaze90}
to test the accuracy of their algorithms.
For computational purposes the array contains four
spheres in the $z$ direction (using six spheres changes
$\sigma_e$ by less than $1\%$).
The values of $(\sigma_e)_M$ for increasing concentration
are plotted along with
the lines of best fit used to estimate $\sigma_e$ in
figure \ref{figsphere}.  The graph demonstrates the necessity
of extrapolating the data to $M^{-1}\to0$.
The results for $\sigma_e$ are presented in
table \ref{tabspheres}. The error is less
than $1\%$ for
$p\leq0.4$ but increases to around $3\%$ at
$p=0.5$ near the percolation threshold $p_c\approx0.52$.
For the random media it was found that computing
$(\sigma_e)_M$ at more than two values of $M$ did
not significantly alter the estimation of $\sigma_e$.

For the random media we must consider how to assign
the conductivity of a bond lying between two nodes
which lie in different phases.
Let $y_{i,j}$ and $\sigma_{i,j}$ be the respective values
of the field and the conductivity at two 
\noindent

\noindent
\begin{minipage}[b!]{8.5cm}
\begin{figure}
\centering \epsfxsize=8.0cm\epsfbox{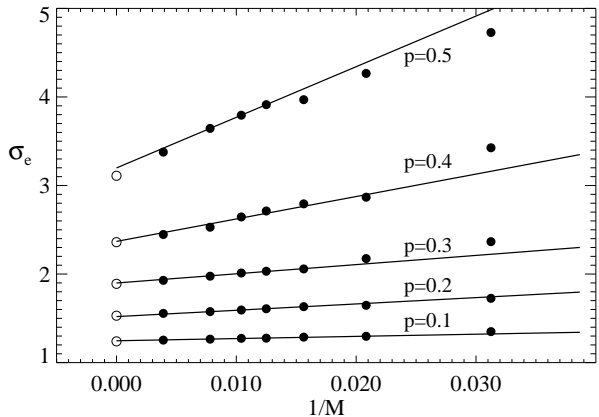}

\vspace{-0mm}

\caption{
\label{figsphere}
The calculated values of $(\sigma_e)_M$(solid circles) for varying
volume fraction ($p$) and the exact data (open circles)
\protect\cite{McKenzie78} for the regular array of spheres.
$M$ is the discretisation used in the finite difference
scheme. 
The lines are linear least squares fits of the
data over the range of $M^{-1}$ where the data is approximately
linear (generally $M^{-1}<0.016$).
$\sigma_e$ is then then given by the intercept of the lines
with the axis $M^{-1}=0$.}
\end{figure}
\end{minipage}

\vspace{2mm}

\noindent
such neighbouring nodes. 
There are three obvious ways of
determining
$\sigma_{ij}$.
Defining $a=(\alpha-y_i)/(y_j-y_i)$ we can choose
$\sigma_{ij}=a \sigma_j + (1-a) \sigma_i$ (as if the portions of
the volume element associated with the bond are like conductors in
{\it parallel}), $\sigma_{ij}=(a/ \sigma_j + (1-a)/\sigma_i)^{-1}$
(as if in {\it series}) or $\sigma_{ij}=\sigma_1$ or $\sigma_2$ as
$(y_i+y_j)/2>\alpha$ or $(y_i+y_j)/2<\alpha$
(a simple field {\it average}).
In figure \ref{figbondrule} we show the effect of using these
rules for two samples of material I generated with $N=4,16$.
For a given discretisation ($M$) a large difference in
$\sigma_e$ occurs depending on the rule employed.
However extrapolation to $M^{-1}=0$ demonstrates
remarkably well that the choice is immaterial.
As the simple averaging rule provides the least error
for given $M$ it will be used.
Finally, for a given volume fraction we use the bisection method
to calculate the value of the level cut parameter $\alpha'$
such that $<\sigma>=p\sigma_1+q\sigma_2$
(where $<>$ refers to bond averaging).
This substantially reduces the statistical fluctuations in 
$\sigma_e$ compared to using the theoretical value of $\alpha$
determined from equation (\ref{onepoint}).

\vspace{-1mm}

\noindent
\begin{minipage}[b!]{8.5cm}
\begin{table}
\caption{Comparison of the extrapolated finite difference
simulations with the exact results \protect\cite{McKenzie78}
for a simple cubic array of spheres of conductivity 
$\sigma_1=10$ in a matrix of conductivity $\sigma_2=1$.
The Brownian motion simulations of
Kim \& Torquato \protect\cite{Kim91} 
and the simulations of Bonnecaze \& Brady
\protect\cite{Bonnecaze90} are also included.
The spheres touch at $p=\pi/6\approx0.52$.}
\label{tabspheres}
\begin{tabular}{cccc|cc}
  & Exact & Finite Diff. & Relative & Kim- & Bonnecaze- \\
p &
Results &
This work &
Error &
Torquato &
Brady \\
\tableline
0.1 &  1.24 & 1.25 & 0.8\% & 1.24 & 1.24 \\
0.2 &  1.53 & 1.52 & 0.7\% & 1.53 & 1.53 \\
0.3 &  1.89 & 1.90 & 0.5\% & 1.89 & 1.87 \\
0.4 &  2.36 & 2.37 & 0.4\% & 2.36 & 2.29 \\
0.5 &  3.11 & 3.19 & 2.6\% & 3.13 & 2.80 
\end{tabular}
\end{table}
\end{minipage}

\noindent
\begin{minipage}[b!]{8.5cm}
\begin{figure}
\centering \epsfxsize=8.0cm\epsfbox{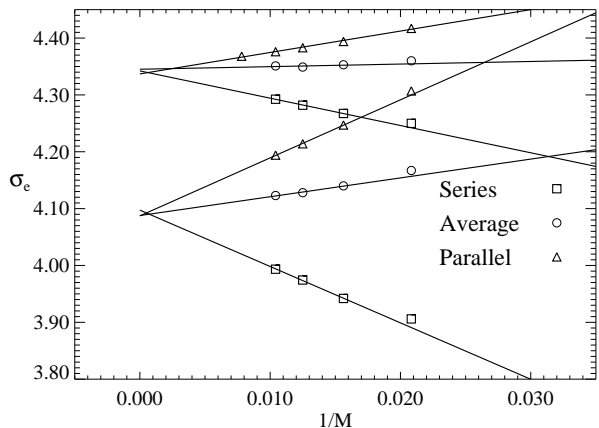}
\caption{
\label{figbondrule}
The values of $(\sigma_e)_M$ for sample Gaussian media
using the three different rules for assigning a conductivity
to each bond (see text).
Note that large variations in $(\sigma_e)_M$ occur for
given $M$ if different rules are employed.
However the choice is seen to be immaterial when the data is
extrapolated to $M^{-1}\to0$.}
\end{figure}
\end{minipage}

\vspace{-8mm}

\section{Simulation Results}
\label{results}

\vspace{-5mm}

We simulate the effective conductivity for the four GRI
models listed in table \ref{tabcomp} for a range compositions.
As we are dealing with finite sized samples we report
$\sigma_e$ as the average over a number of different realizations
of the materials. 
Error bars, which represent 95\% confidence limits on
the results, are equal to twice the standard error. 
The samples are examined at three different contrast values;
$\sigma_{1,2}=10,1$, $\sigma_{1,2}=50,1$ and $\sigma_{1,2}=1,0$.
Previous authors \cite{Kim91,Kim92,Bonnecaze91}
have considered media with $\sigma_{1,2}=\infty,1$
but this is not possible using the methods discussed here.

The results for the case $\sigma_{1,2}=10,1$ are presented
in table \ref{data10_1} and plotted in figure \ref{figGRI10-1}
for $p\in[0.1,0.9]$.
To obtain the data five samples of each model with
discretisations using $M=64$ and $M=96$ were considered.
The results show little variation in $\sigma_e$ (for fixed $p$)
amongst the four different media.
The maximum relative difference of $4.2\%$ occurs at
$p=0.6$ between $\sigma_e$ for model I ($\nu=0$, $K=8$)
and model III ($\mu=1.5$).
As the differences are relatively small we restrict further
attention to the latter two materials.
The bounds calculated from equations
(\ref{BMhi}) and (\ref{Mlo}) using $\zeta_1$ from
tables \ref{tabzeta1} and \ref{tabzeta2} are presented
in table \ref{tabbounds} and plotted along
with the simulation data in figures \ref{figC10-1},
for $p\in [0.2,0.8]$, and figure \ref{figK10-1} for
$p\in[0.86,0.96]$.
The latter figure illustrates very clearly that the
upper bound
discriminates between model I and III.

For the case $\sigma_{1,2}=50,1$ the results for model I
($\nu=0$, $K=8$) and model III ($\mu=1.5$) are reported
in table \ref{data50_1} and plotted along with relevant
bounds in figure \ref{figO50-1}.
Again $5$ samples of the media were considered with $M=64$
and $M=96$.
Figure \ref{figO50-1} shows very pronounced differences
between the IOS model and GRI models.
The results for the case $\sigma_{1,2}=1,0$ are given
in table \ref{data1_0} and plotted along with the
upper bound (the lower bound vanishes) in figure \ref{figN1-0}.
Five samples at discretisations $M=48$ and $M=64$
were considered. 

\noindent
\begin{minipage}[b!]{8.5cm}
\begin{table}
\caption{Simulations of $\sigma_e$ for different GRI
models with $\sigma_{1,2}=10,1$.
The data for the IOS model \protect\cite{Kim92}
is included for purposes of comparison.
The notation $x(y)$ implies
$\sigma_e=x\pm y.10^{-2}$ with the error
bars defined as 95\% confidence 
limits \protect\tablenote{$y=0$ implies $y<0.5$}.}
\label{data10_1}
\begin{tabular}{cccccc}
  & I $\nu=0$  & I $\nu=10$ & II & III $\mu=1.5$ & IOS \\
p & K=8         & K=32       &K=8 &               &     \\
\tableline
0.1 &  1.36(1) & 1.33(1) & 1.35(1) & 1.33(1) &      \\
0.2 &  1.84(2) & 1.80(2) & 1.85(2) & 1.82(3) & 1.64 \\
0.3 &  2.45(4) & 2.43(2) & 2.49(2) & 2.47(3) &      \\
0.4 &  3.19(5) & 3.21(3) & 3.24(3) & 3.28(4) & 2.73 \\
0.5 &  4.06(6) & 4.12(3) & 4.12(2) & 4.21(4) &      \\
0.6 &  5.03(6) & 5.12(2) & 5.10(2) & 5.24(6) & 4.63 \\
0.7 &  6.12(6) & 6.23(1) & 6.20(1) & 6.33(6) &      \\
0.8 &  7.33(4) & 7.41(1) & 7.39(1) & 7.50(4) & 7.11 \\
0.86&  8.13(3) &         &         & 8.20(4) &      \\
0.88&  8.39(3) &         &         & 8.46(4) &      \\
0.90&  8.65(2) & 8.68(0) & 8.67(2) & 8.71(3) &      \\
0.92&  8.91(2) &         &         & 8.97(3) &      \\
0.94&  9.18(2) &         &         & 9.22(2) &      \\
0.96&  9.45(1) &         &         & 9.48(1) &
\end{tabular}
\end{table}
\end{minipage}

\noindent
\begin{minipage}[b!]{8.5cm}
\begin{figure}
\centering \epsfxsize=8.0cm\epsfbox{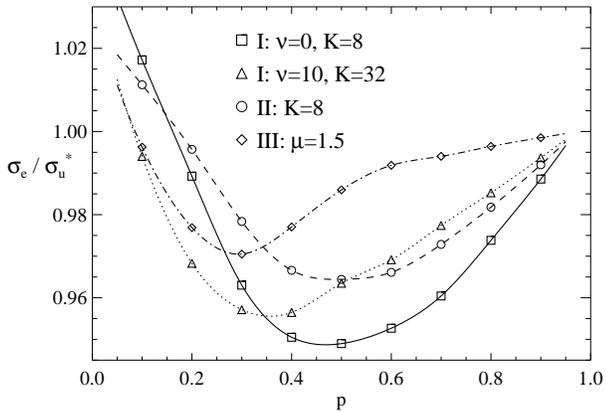}
\caption{
\label{figGRI10-1}
The data for four different GRI models with conductivities
$\sigma_1=10$ and $\sigma_1=1$. The data is normalized
against the BM upper bound (eqn.\ (\protect\ref{BMhi}))
with $\zeta_1=p$ to highlight the differences between the models.
The curves are spline fits of the data.}
\end{figure}
\end{minipage}

\noindent
\begin{minipage}[b!]{8.5cm}
\begin{table}
\caption{An example of the bounds calculated using
equations (\protect\ref{BMhi}) \& (\protect\ref{Mlo})
and the data in tables \protect\ref{tabzeta2} (Model I) \&
\protect\ref{tabzeta1} (Model III). The conductivity contrast
is $\sigma_{1,2}=10,1$}
\label{tabbounds}
\begin{tabular}{cccccccc}
\multicolumn{1}{c}{ } &
\multicolumn{2}{c}{I $K$=8, $\nu=0$}   &
\multicolumn{2}{c}{III $\mu=1.5$}   \\
\hline
p & $\sigma_l$ & $\sigma_u$ & $\sigma_l$ & $\sigma_u$ \\
\tableline
0.1  & 1.290 & 1.437 & 1.268 & 1.372  \\
0.2  & 1.660 & 1.994 & 1.618 & 1.905  \\
0.3  & 2.123 & 2.654 & 2.073 & 2.576  \\
0.4  & 2.702 & 3.414 & 2.663 & 3.371  \\
0.5  & 3.423 & 4.273 & 3.423 & 4.273  \\
0.6  & 4.316 & 5.229 & 4.385 & 5.269  \\
0.7  & 5.414 & 6.283 & 5.570 & 6.348  \\
0.8  & 6.742 & 7.434 & 6.961 & 7.501  \\
0.9  & 8.302 & 8.677 & 8.490 & 8.720 
\end{tabular}
\end{table}
\end{minipage}

\vspace{-9mm}

\noindent
\begin{minipage}[b!]{8.5cm}
\begin{figure}
\centering \epsfxsize=8.0cm\epsfbox{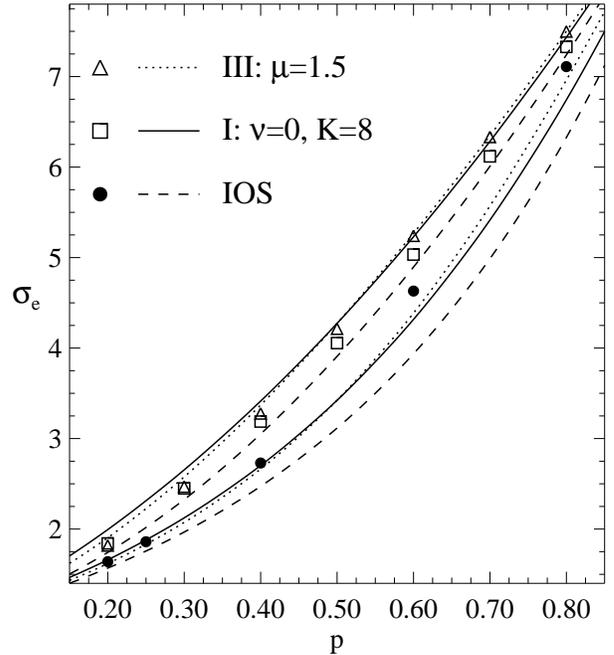}
\caption{
\label{figC10-1}
Simulations and bounds for the effective conductivity of
two models of random composite media generated from the
GRI model for the case $\sigma_{1,2}=10,1$. 
The circular symbols represent data for the IOS model
calculated by Kim \& Torquato \protect\cite{Kim92}
and the bounds for the IOS model were evaluated by
Torquato \& Stell \protect\cite{TorqStel83}.}
\end{figure}
\end{minipage}

\end{multicols}

\vspace{-6mm}

\begin{multicols}{2}

\noindent
\begin{minipage}[b!]{8.4cm}
\begin{figure}
\centering \epsfxsize=8.0cm\epsfbox{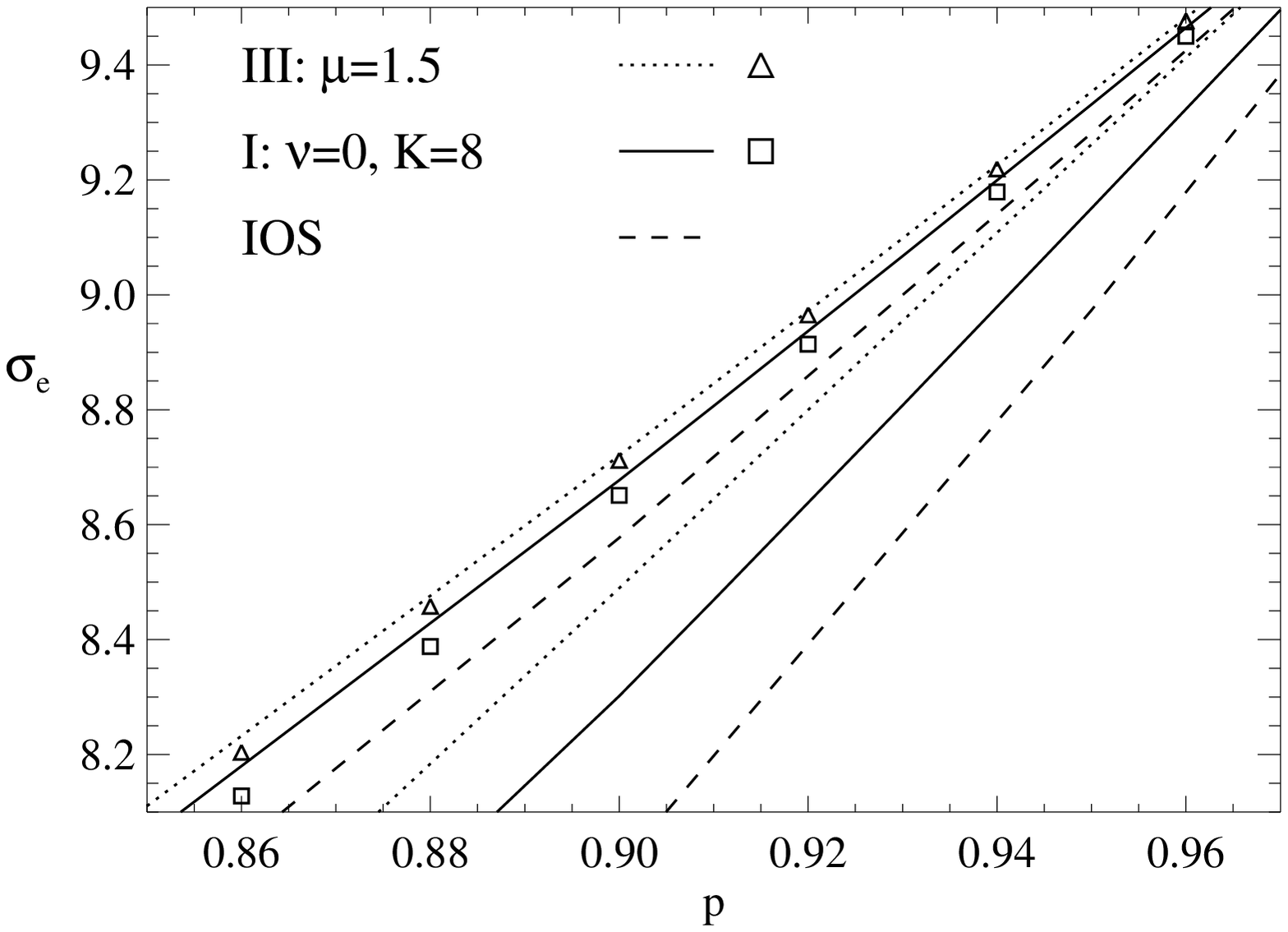}
\end{figure}
\end{minipage}

$\;$

\vspace{12mm}

$\;$

\noindent
\begin{minipage}[b!]{8.4cm}
\begin{figure}
\caption{
\label{figK10-1}
Simulations and bounds for the effective conductivity of
two models of random composite media generated from the
GRI model for the case $\sigma_{1,2}=10,1$.
The upper bound discriminates between the models
in this range of $p$.}
\end{figure}
\end{minipage}
\end{multicols}

\clearpage

\begin{multicols}{2}

\noindent
\begin{minipage}[b!]{8.5cm}
\begin{figure}
\centering \epsfxsize=8.0cm\epsfbox{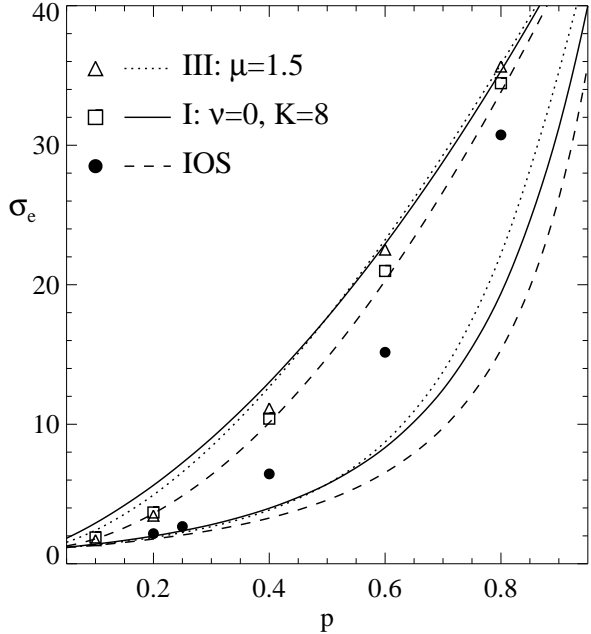}
\caption{
\label{figO50-1}
As in figure \protect\ref{figC10-1} except $\sigma_1=50$ and
$\sigma_2=1$.}
\end{figure}
\end{minipage}
\noindent
\begin{minipage}[b!]{8.5cm}
\begin{table}
\caption{Simulations of $\sigma_e$ for the different
Gaussian random media with $\sigma_1=50$ and $\sigma_2=1$.
The IOS model data is from ref.\
\protect\cite{Kim92}.
The error bars define 95\% confidence limits.}
\label{data50_1}
\begin{tabular}{cr@{$\pm$}lr@{$\pm$}lc}
\multicolumn{1}{c}{ } &
\multicolumn{2}{c}{I $\nu=0,K=8$} &
\multicolumn{2}{c}{III $\mu=1.5$} &
\multicolumn{1}{c}{IOS} \\
\tableline
0.1 &  1.88 & 0.08 & 1.69 & 0.06 &           \\
0.2 &  3.68 & 0.2  & 3.45 & 0.2  &  2.16     \\
0.4 &  10.4 & 0.5  & 11.1 & 0.3  &  6.44     \\
0.6 &  21.0 & 0.4  & 22.5 & 0.3  &  15.2     \\
0.8 &  34.4 & 0.4  & 35.6 & 0.2  &  30.7    
\end{tabular}
\end{table}
\end{minipage}
\begin{minipage}[b!]{8.5cm}
\begin{figure}
\centering \epsfxsize=8.0cm\epsfbox{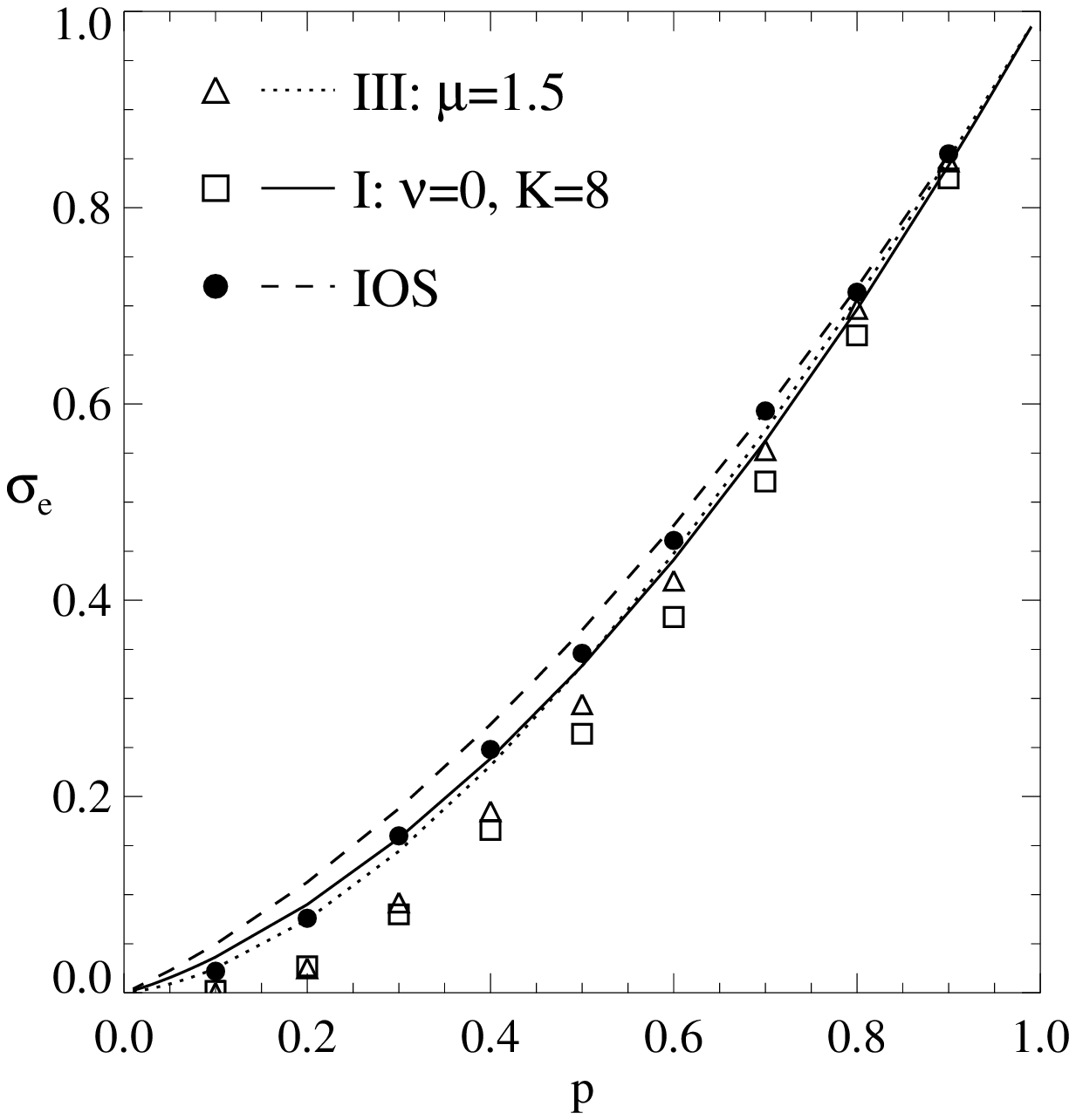}
\caption{
\label{figN1-0}
As in figure \protect\ref{figC10-1} except $\sigma_1=1$ and
$\sigma_2=0$.}
\end{figure}
\end{minipage}
\begin{minipage}[b!]{8.5cm}
\begin{table}
\caption{Simulation of $\sigma_e$ for different Gaussian
random media with $\sigma_1=1$ and $\sigma_2=0$.
The IOS model data is from ref.\ \protect\cite{Kim92}.
The notation $x(y)$ implies
$\sigma_e=x\pm y.10^{-3}$ with the error
bars defined as 95\% confidence
limits \protect\tablenote{$y=0$ implies $y<0.5$.}.}
\label{data1_0}
\begin{tabular}{llll}
p & I, $\nu=0,K=8$ & III $\mu=1.5$ & IOS \\
\tableline
0.1 &  0.002(2)  & 0.000(0)  & 0.022\\
0.2 &  0.027(6)  & 0.025(2)  & 0.076\\
0.3 &  0.080(3)  & 0.092(6)  & 0.160\\
0.4 &  0.166(12) & 0.185(10) & 0.248\\
0.5 &  0.264(10) & 0.294(12) & 0.346\\
0.6 &  0.383(9)  & 0.420(11) & 0.461\\
0.7 &  0.521(12) & 0.553(6)  & 0.593\\
0.8 &  0.670(7)  & 0.697(7)  & 0.714\\
0.9 &  0.830(3)  & 0.847(2)  & 0.855
\end{tabular}
\end{table}
\end{minipage}
\end{multicols}
\begin{multicols}{2}

There are several qualitative trends in the data
which can be commented on.
Note that for first two contrasts considered $\sigma_e$
of the GRI models is greater than that of the IOS model
over the entire range of $p$.
At low volume fractions this can be attributed to the
fact that the inclusions of the GRI models are qualitatively
less spherical than those of the IOS model (see appendix
\ref{qualitative}).
This can be clearly seen in figures \ref{figIp.07}
and \ref{figIIIp.07} where the inclusions are plotted for
each of the GRI models at $p=0.07$ (the IOS model will
containing predominantly spherical inclusions at this
volume fraction).
At high volume fractions the situation is reversed;
the matrix phase of the IOS model is extremely ramified
and hence $\sigma_e$ is lower.
Similarly near $p=0,1$ the small differences in
$\sigma_e$ for Models I and III can be explained by the fact
that the latter model has more spherical inclusions
(compare figures \ref{figIp.07} \& \ref{figIIIp.07}).
This behavior is consistent with the relative
variations in $\zeta_1$ for each of the three models as
discussed in appendix \ref{qualitative}.
For mid-range $p$ the differences between the IOS and GRI models
correspond to the fact that the more
highly conducting regions of the latter are generally better
connected than those of the IOS model. Again this difference can be 
anticipated from the relative behavior of the parameter $\zeta_1$
for the two classes of models.
However this is not necessarily always so
as can be seen by comparing the respective values of
$\zeta_1$ (tables \ref{tabzeta1} \& \ref{tabzeta2}) and
$\sigma_e$ (table \ref{data10_1}) for Models I and III at $p=0.4$.
In this case $\sigma_e^{\rm I}<\sigma_e^{\rm III}$ but
$\zeta_1^{\rm I} > \zeta_1^{\rm III}$.

\end{multicols}
\begin{multicols}{2}

\noindent
\begin{minipage}[b!]{8.5cm}
\begin{figure}
\centering \epsfxsize=7.0cm\epsfbox{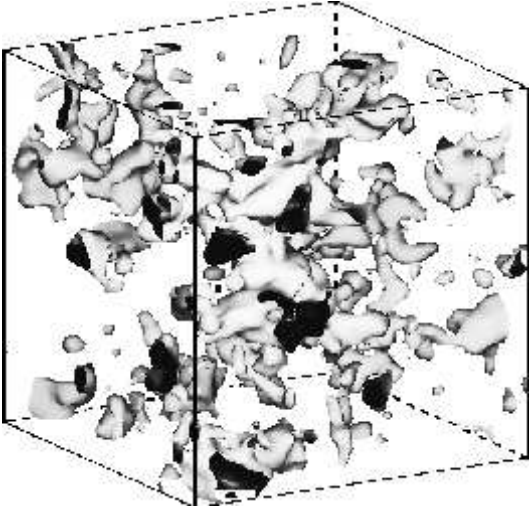}

\vspace{3mm}

\caption{
\label{figIp.07}
A plot of the interface
$y_{\!\!\!\mbox{ {\protect\tiny $K$} }\!\!\!} ({\bf r})=1.48$
($p=0.07$) for Model I.
The parameters used to generate the field are $\nu=0,K=8,T=4\pi$.}
\end{figure}
\end{minipage}
\noindent
\begin{minipage}[b!]{8.5cm}
\begin{figure}
\centering \epsfxsize=7.0cm\epsfbox{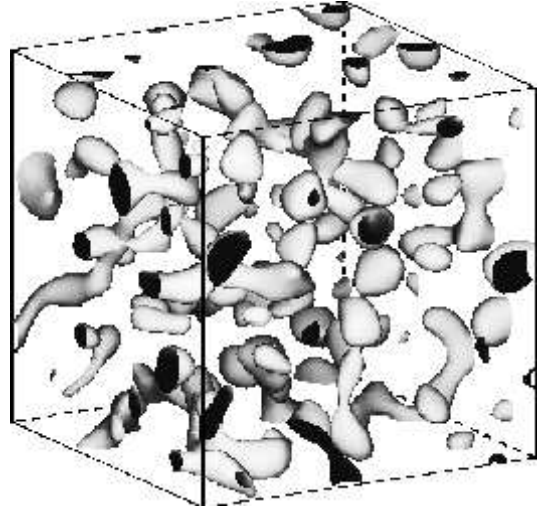}

\vspace{3mm}

\caption{
\label{figIIIp.07}
A plot of the interface
$y_{\!\!\!\mbox{ {\protect\tiny $K$} }\!\!\!} ({\bf r})=1.48$
($p=0.07$) for Model III.  The parameters used to generate
the field are $\mu=1.5,T=8\pi$.}
\end{figure}
\end{minipage}
\end{multicols}
\begin{multicols}{2}

Note that the simulation data for the IOS model in the case
$\sigma_{1,2}=1,0$ were obtained for
insulating spheres in a matrix of unit conductivity.
Therefore the microstructure of the conducting phase is
generally better connected than the GRI models and the
arguments pertaining to the qualitative variations in
$\sigma_e$ used above are reversed.

In all of the above cases the data lies between the
appropriate bounds.
Furthermore the upper bound is seen to provide a good estimate for
$\sigma_e$ for the GRI models for $p\geq0.7$.
This has been observed previously
\cite{TorqRev91,Torquato85a,Kim92} for both high and low $p$
(the latter case has not been considered in detail here).
This fact provides evidence of the near optimality of bounds
in these regions and a good confirmation of the techniques used
in this paper (see figure \ref{figK10-1} especially). 

We have calculated the percolation threshold for
Model I and III using the algorithm of Skal {\it et al.}
\cite{Skal73a}.
In contrast to previous results \cite{Skal73b} we found
$p_c$ to exhibit finite
size effects and to depend on the spectrum.
Averaging $p_c$ over 10 fields at $M=20,32,48,64$
and extrapolating
to $M=\infty$ we found
$p_c\approx 0.07$ ($\alpha_c\approx1.47$) for Model I
and $p_c\approx0.13$ for Model III ($\alpha_c\approx1.13$).
In the absence of theoretical percolation results for
the GRI model it is interesting to discuss the threshold
in terms of the transition of the structures from elliptic
to hyperbolic as $p$ increases.
The average Gaussian curvature ($K_G$) for the GRI models is given
by \cite{Teubner91},
\begin{equation}  <K_G>=\frac16 <k^2> (\alpha^2-1).  \end{equation}
Therefore the nature of the interface is dominated
by inclusions of positive curvature
(eg. ellipsoids)
for $|\alpha|>1$ and is predominantly hyperbolic
(eg. bicontinuos) for $|\alpha|<1$.
The fact that $\alpha_c > 1$ indicates that connecting
structures persist below the elliptic/hyperbolic transition
as would be expected since $<K_G>$ is an average quantity.

Finally it is interesting to discuss the surprisingly small
variation in $\sigma_e$ (and $\zeta_1$) amongst
the GRI materials.
As
 can be seen in figures \ref{figmodels} (a)-(d)
these materials appear to be very different when viewed
at the same scale.
However the major qualitative differences are related
to the effective decay length, and when the materials are
viewed at a scale proportional to this length they
appear to be remarkably similar as shown in
figures \ref{figmodels} (e)-(h). 
Equivalently if the length parameters are retained in eqns.\
(\ref{moda})-(\ref{modf}) they can be tuned to achieve
the latter group of figures at the same scale
(without effecting $\sigma_e$).
The remaining smaller qualitative differences account
for the variation in $\sigma_e$ observed.

\section{Conclusion}
\label{conclusion}
We have investigated the effective conductivity of a
two phase random composite material using bounding techniques
and direct simulation.
Our calculations of $\zeta_1$ increase the classes of composites to
which the bounds can be applied, while the simulation data can be used
to assess predictive theories~\cite{RobertsUP1} and be compared with
higher order bounds.
The results also pertain to a variety of other
effective properties of amorphous composites as discussed
in section \ref{bounds}.

The bounds encompass all of the simulational data and
the upper bound yields a reasonable estimate of
$\sigma_e$ for $p\geq 0.7$ ($\sigma_1 > \sigma_2$). 
Reasonably large differences in $\sigma_e$ and
$\zeta_1$ are observed between the amorphous GRI models
and the IOS model.
This highlights the importance of incorporating microstructure
effects in the calculation of the effective properties of
composite materials.
Conversely there is relatively little variation
in $\sigma_e$ amongst the GRI models as the major qualitative
microstructural differences between the models are related to an
effective decay length upon which $\sigma_e$
is necessarily independent.
We expect that other properties of such composites
where no intrinsic field length scale is present
(eg. the elastic bulk and shear moduli)
will show similar behavior. 

It is clear that the Gaussian random interface model
discussed here can serve as a useful `model' amorphous medium
in the study of the effective properties of random composites.
Furthermore the bounds and simulations can be related to physical
composites by experimentally relating such systems to one of the
theoretically known models.
This could be done by comparing the spectra obtained from small
angle scattering studies with that obtained from the 2-point
correlation functions of each model. Or, more simply,
by comparing images of the models with electron micrographs.
Although such schemes are only approximate, our results indicate
that the fine microstructural details are relatively unimportant
in both the calculation of $\zeta_1$ and the simulation of $\sigma_e$.

We note that the GRI model can be extended to the
case of membranes and foams \cite{Berk87,RobertsUP1} and that higher order
correlation functions can be calculated for use in more precise bounds.
Random walker algorithms can be utilised to investigate the
often studied scaling properties of $\sigma_e$ near
the percolation threshold.

\acknowledgments
The authors thank Stjepan Mar\v{c}elja for suggesting the
problem and M.  Knackstedt, K. McGrath, P. Pieruschka,
D. Singleton and X. Zhang for helpful discussions.
The simulations reported here were carried out on the
{\it Thinking Machines} CM-5 at the Australian
National University Supercomputer Facility.
A.R. is supported by an Australian postgraduate research award.  
\appendix
\section{Asymptotic forms of $p_2$}
\label{asymptotic}
The two point function (eqn.\ (\ref{twopoint})) can be
expanded in powers of $\alpha$ to yield,
\begin{equation}  p_2(g)= \frac1{2\pi} \arcsin g +
\frac{ e^{-\frac12\alpha^2}}{2\pi}
\sum_{n=1}^\infty
\frac{(-1)^n \alpha^{2n} a_n }{2^n n!}+p^2
\end{equation}
with,
\begin{equation}
a_n=\frac 2{2n-1} 
\left( 1 -\left( \frac{1-g}{1+g} \right)^{n-\frac12}
\right)-a_{n-1}
\end{equation}
and $a_0=0$.
This expansion converges rapidly for small $|\alpha|$.

For the case $\alpha \gg 0$ we have
(using successive integration by parts),
\begin{equation}
\label{asy21} p\sim\frac{e^{-\frac12\alpha^2}}{\sqrt{2\pi}\alpha}
\left( 1+\sum_{n=1}^{\infty}\frac{(-1)^n1.3\dots(2n-1)}{\alpha^{2n}}
\right).  \end{equation}
Special care must be taken to determine a practical
expansion for $p_2$.
A simple application of Watson's Lemma\cite{Murray}
yields a solution which does not possess the correct
limiting behavior as $g\to0$ and gives two different
expansions for the cases $g=1$ and $g<1$.
The first problem is dealt with by appropriately partitioning
the integral, while the second necessitates a further
transformation of variable, followed by an expansion in scaled
parabolic cylinder functions 
(see for example ref.\ \cite{Bleistein66}).  
Thus we write
\begin{equation}
\label{asy22}  p_2(g)= \frac1{2\pi}
\left( \int_{-1}^g\!\!-\!\!\int^{0}_{-1} \right){
\exp{\left(-\frac{\alpha^2}{1+t}\right)}
{\frac{dt}{\sqrt{1-t^2}}
}}+p^2.
\end{equation}
In the second of these integrals we make the substitution
$v=1/(1+t)-1$ to give
\begin{equation}
\frac{e^{-\alpha^2}}{2^\frac32\pi}\int_{0}^{\infty}
{\frac{e^{-\alpha^2 v}}{(v+1)(v+\frac12)^\frac12}}{dv}
\end{equation}
which can be expanded using Watson's Lemma,
\begin{equation}
\frac{e^{-\alpha^2}}{2\pi}
\left( {\frac1{\alpha^2}-\frac2{\alpha^4}+\frac7{\alpha^6}+O
\left( {\frac1{\alpha^8} }\right) }\right).
\end{equation}
This is just the expansion of $p^2$ which can be
cancelled from equation~(\ref{asy22}).
The remaining integral is put in a standard form by
the substitution $v=1/(1+t)-1/(1+g)$,
\begin{equation}
p_2(g)=
\frac{e^{-\frac{\alpha^2}{1+g}}}{2^\frac32\pi}\int_{0}^{\infty}
{\frac{e^{-\alpha^2
v}}{(v+\frac1{1+g})(v+\frac12\frac{1-g}{1+g})^\frac12}}{dv}.
\end{equation}
Note that the nature of the singularity of the
integrand changes order as $g\to1$
(this makes it impossible to generate an expansion for the
full range of $g$ using Watson's Lemma).
To develop a uniform expansion for large $\alpha$ valid near
$g=1$ we make the further substitution,
\begin{equation}
v=\frac12u^2+\delta_g u,\;\;\;
\delta_g=\sqrt\frac{1-g}{1+g}, 
\end{equation}
to give
\begin{equation}
\frac{e^{-\frac{\alpha^2}{1+g}}}{2\pi}\int_{0}^{\infty}
{\frac{e^{-\alpha^2(\frac12 u^2+\delta_g u)}}
{\left( { \frac12 u^2+\delta_g u +\frac{1}{1+g} }\right) } }{du}. 
\end{equation}
In the usual way the non-exponential component of the
integrand can be expanded in powers of $u$ and
integrated term by term to give,
\begin{eqnarray}
p_2(g)&\approx&
\frac{e^{-\frac{\alpha^2}{1+g}}}{2\pi} \left(
\frac{1+g}\alpha T_1\left( {\delta_g\alpha} \right)
-\frac{(1+g)^2}{\alpha^2} \delta_g
 T_2\left( {\delta_g\alpha} \right) \right. \nonumber \\
&+& \left. \frac{(1-2g)(1+g)^2}{2\alpha^3}T_3
\left( {\delta_g\alpha} \right) \right) 
\label{asy28}
\end{eqnarray}
where,
\begin{equation}
T_n(z)=\int_{0}^{\infty}\!\!\!{e^{-zs-\frac12s^2}s^{n-1}}{ds}
=\Gamma(n)e^ {\frac14{z^2}}{\rm U}(n\!-\!\frac12,z). 
\end{equation}
${\rm U}(a,z)$ is a parabolic cylinder function\cite{Abramowitz}.
Two simple checks can be made on this expansion.
For $g=1$ we have $ T_n(0)=2^{n/2 -1}\Gamma(\frac n2)$  which,
when substituted in (\ref{asy28}), gives the expansion of $p$.
For $g<1$ we again employ Watson's Lemma to determine
the asymptotic expansion,
\begin{equation}
T_n(z)=\sum_{j=0}^\infty\frac{(-1)^j\Gamma(n+2j)}{j! 2^j z^{n+2j}}.
\end{equation}
Now taking $g=0$ in (\ref{asy28}) and using this expansion gives
the asymptotic form of $p^2$ as it should.
The expansions for the case $\alpha \ll 0$ ($p\approx 1$)
are simply, $p=1-{\rm AE}_1$ and $p_2(g)=2p-1+{\rm AE}_2$
where ${\rm AE}_1$ and ${\rm AE}_2$ are the asymptotic expansions
given by (\ref{asy21}) and (\ref{asy28}) respectively
with $\alpha$ replaced by $|\alpha|$.
Note that Berk\cite{Berk87,Berk91} has derived a formal
series representation of $p_2$ valid for all $\alpha$,
however the convergence of the series is slow for $g\approx1$
and not guaranteed at $g=1$.
\section{Proof of $\zeta_1=\frac12$ for $p=\frac12$.}
\label{zhalf}
In Brown's formulation the parameter $\zeta_1$ arises
as the limit as $\epsilon\to0$ of the integral,
\begin{equation}
\label{proofzeta}
\zeta_1=
\frac9{2pq}\int_\epsilon^\infty\!\!\frac{dr}{r}
\int_\epsilon^\infty\!\! \frac{ds}{s} 
\int_{-1}^1 du  P_2(u) f(r,s,t) 
\end{equation}
where $f$ is the term in brackets of eqn.\ (\ref{zeta}).
Now by taking $p=\frac12$ in eqn.\ (\ref{threepoint}) we have
$f=p_2^T(t)/2-2p_2^T(r)p_2^T(s)$.
Note that the integral of the second
term vanishes since it does not depend on $u$
and $\int_{-1}^1P_2(u)du=0$.
Therefore after making
the substitution $t^2=r^2+s^2-2sru$ eqn.\ (\ref{proofzeta})
becomes,
\begin{equation}
\zeta_1=9\!\int_\epsilon^\infty\!dr \int_\epsilon^r\! ds
\int_{r-s}^{r+s} p_2^T(t) h(r,s,t) dt 
\end{equation}
where $h(r,s,t)=t ( s r )^{-4} (\frac34(t^2-s^2-r^2)^2- r^2 t^2 )$.
Interchanging the order of integration results in,
\begin{eqnarray} \nonumber
\zeta_1&=&9\left( \int_0^{2\epsilon}+
\int_{2\epsilon}^{\infty} \right) p_2^T(t) 
\int_{\epsilon}^{\infty} ds \int_{s}^{t+s} h dr \\
&-&9\int_{2\epsilon}^{\infty} p_2^T(t) \int_{\epsilon}^{t/2} ds
\int_{s}^{t-s} h dr,
\end{eqnarray}
and carrying out the straight forward integrations over
$r$ and $s$ leads to,
\begin{equation}
\zeta_1 = 9\! \int_0^{2\epsilon} p_2^T(t) 
\left(\frac{2t^2}{3\epsilon^3}-\frac{t^3}{2\epsilon^4}
+\frac{t^5}{24\epsilon^6}\right)dt.
\end{equation}
Now by taking $\epsilon\to0$ and using the fact that
$p_2^T(0)=\frac14$ gives $\zeta_1=\frac12$. 
\section{Relationship between $\sigma_e$,
$\zeta_1$ and shape at low $p$ }
\label{qualitative}
Consider the small concentration approximation \cite{Reynolds57}
to $\sigma_e$ for the case of randomly distributed and
oriented spheroids with axial ratios $A,A,1-2A$,
\begin{equation}  \sigma_e=\sigma_2+\frac13 p(\sigma_1-\sigma_2)
z(\sigma_1,\sigma_2,A)+O(p^2) \label{spheroids}
\end{equation}
where,
\begin{equation}
z=\frac{2\sigma_2}{\sigma_2+A(\sigma_1-\sigma_2)}+
\frac{\sigma_2}{\sigma_2+(1-2A)(\sigma_1-\sigma_2)}.
\end{equation}
With $A\in[0,1/2]$ it can be easily shown that $z$
has a unique minimum at $A=1/3$ (spherical inclusions)
and is monotonically increasing as $|A-1/3|$ increases.
Therefore with $\sigma_1>\sigma_2$, $\sigma_e$
will be higher the lower the sphericity of the
inclusions (and conversely for $\sigma_2>\sigma_1$).
The same argument should qualitatively hold
for arbitrary shapes.
To see how this relates to $\zeta_1$ we match
the terms of the expansions of eqn.\ (\ref{spheroids})
and eqn.\ (\ref{BMhi}) to order $p$ and
$(\sigma_1-\sigma_2)^3$ which gives \cite{Miller69,Torquato85b}
$\zeta_1=(1-3A)^2 +O(p)$.
Thus $\zeta_1$ will be higher for less spherically shaped
inclusions.
\section{Implementation of the finite difference scheme}
\label{implementation}
The finite difference scheme (see for example ref.\ \cite{Press})
for the equations and boundary conditions discussed in
section \ref{bounds} leads to a system of simultaneous equations
for the value of the potential at each of the interior nodes
(including those on lateral faces if we define $\phi$ to be
periodic in the $x$ and $y$ directions).
For each such node $u$ we have
\begin{equation}
\sum_{v \in nn} \sigma_{uv}(\phi_{u}-\phi_{v}) =0 \label{fds}
\end{equation}
where $nn$ is the set of nearest neigbours of node 
$u$ and $\sigma_{uv}$ is the conductivity of the bond
lying between nodes $u$ and $v$.
Conventionally these equations are cast as a
matrix equation with $\phi$ and the boundary conditions
as 1 dimensional matrices (vectors) \cite{Press}.
On a parallel computer it saves coding and implementation
time to retain the potential $\phi_{i,j,k}$
as a 3D matrix.
Define $A$ as the operator which performs the operations
defined on the left hand side of (\ref{fds}) for interior
nodes and $A\phi\equiv\phi$ on the nodes where Dirichlet
conditions are to be applied.
Also define $b$ as a 3D matrix containing the boundary
conditions on the field ($b_{i,j,1}=\phi_1,\; b_{i,j,M}=\phi_0$
for all $i,j$ and $b=0$ elsewhere).
Then solving the system of equations for $\phi$ is
equivalent to minimising $ \parallel A \phi - b \parallel_2$, 
which can be done using a conjugate gradient method
which handles vectors of general dimension.
\section{Calculation of the microstructure parameter $\eta_1$}
\label{elastic}
Three point bounds have also been derived for the
elastic bulk and shear moduli
\cite{Beran65b,McCoy,Milton81a,Milton82b}
which can be expressed \cite{Milton81a} in terms of
$\zeta_1$ and an additional parameter,
\begin{eqnarray} \nonumber
\eta_1&=&\frac 5{21}\zeta_1+
\frac{150}{7pq}\int_0^\infty\!\!\frac{dr}{r}
\int_0^\infty\!\! \frac{ds}{s}
\int_{-1}^1 du  P_4(u) \times \\
&&\left( p_3(r,s,t)-\frac{p_2(r) p_2(s)}{p}
\right).
\label{eta}
\end{eqnarray}
We have calculated $\eta_1$ for several different GRI models and
tabulated (see table \ref{tabeta}) it along with data for the
IOS model.
Qualitatively the results are similar to those for
$\zeta_1$ and we expect the differences in the effective shear
and bulk moduli amongst the GRI models to be similar to those
observed for the conductivity case.

\noindent
\begin{minipage}[b!]{8.5cm}
\begin{table}
\caption{The microstructure parameter $\eta_1$ which
arises in bounds on the elastic bulk and shear moduli 
(see appendix \protect\ref{elastic})
for selected Gaussian media. Data for the IOS model 
\protect\cite{TorqRev91,TorqStel85b} is included
for purposes of comparison.}
\label{tabeta}
\begin{tabular}{ccccc}
  & IOS & I $\nu=0$ & I $\nu=0$  & III $\mu=1.5$ \\
p &     & $K=\infty$ & $K=8$       &               \\
\tableline
0.1  & 0.075 & 0.276 & 0.213 & 0.106    \\
0.2  & 0.149 & 0.333 & 0.291 & 0.197    \\
0.3  & 0.224 & 0.388 & 0.362 & 0.294    \\
0.4  & 0.295 & 0.444 & 0.432 & 0.396    \\
0.5  & 0.367 & 0.500 & 0.500 & 0.500    \\
0.6  & 0.439 & 0.556 & 0.568 & 0.604    \\
0.7  & 0.512 & 0.612 & 0.637 & 0.706    \\
0.8  & 0.583 & 0.667 & 0.709 & 0.803    \\
0.9  & 0.658 & 0.724 & 0.787 & 0.894       
\end{tabular}
\end{table}
\end{minipage}

\end{multicols}

\end{document}